\documentclass{ws-procs975x65}

\begin{document}

\title{Approximate waveform templates for detection of extreme mass ratio inspirals
with LISA}

\author{JONATHAN R GAIR}

\address{Institute of Astronomy, University of Cambridge, Cambridge, CB3 0HA, UK\\
\email{jgair@ast.cam.ac.uk}}

% WARNING. in standard latex cls file formatting, at this point
% \maketitle would typeset the above titlepage information
% but WS has chosen to be nonstandard and have each line typeset 
% as it is digested.
% no abstract is necessary.
% \bodymatter below resets the footnote counter and symbols after 
% possible use in the title matter.

\bodymatter

\section{Introduction}\label{intro}
One of the most interesting potential sources of low frequency gravitational waves (GWs) for LISA are the inspirals of stellar mass compact objects (white dwarfs, neutron stars or black holes) into supermassive black holes (SMBHs) in the centers of galaxies. The mass ratio is typically $10:10^6$, so these events are termed extreme mass ratio inspirals (EMRIs). Detection and parameter estimation for these events is likely to involve matched filtering. LISA will observe EMRIs for the last several years of inspiral prior to plunge, so the search templates will need to match the phase of the signal over a few$\times 10^5$ cycles. The extreme mass ratio ensures that templates of sufficient accuracy can be computed using black hole perturbation theory --- the ``self-force'' formalism \cite{poissonLR}. Such templates are not yet available, however, and will be very computationally expensive when they are ready. The number of templates required to cover the parameter space of possible EMRI signals is very large \cite{gair04}, so there is a need for approximate models that are quick to generate while also being able to estimate the parameters of the source with sufficient accuracy that follow-up with more accurate waveforms is possible.

One family of models are ``adiabatic'' templates, which are based on accurate evaluation of the dissipative part of the self-force, combined with the assumption that the orbital inspiral occurs slowly compared to the orbital period \cite{hughes05,drasco06}. Adiabatic waveforms are likely to play a role at some stage of the LISA data analysis pipeline, but they are still computationally expensive. For scoping out LISA data analysis, waveforms must be generated in large numbers, so two families of approximate, quick-to-compute, ``kludge'' waveforms have also been developed. The ``analytic kludge'' is a phenomenological model based on Keplerian waveforms with relativistic inspiral and precession imposed \cite{BC04}. The ``numerical kludge'' (NK) will be described here. These NK waveforms are sufficiently faithful that they may play a role in source detection for LISA and perhaps in source characterization.

\section{Numerical Kludge Waveform Model}
The NK family of waveforms are designed to be faithful models of true EMRI GW signals. The waveform parameters are the same as for true EMRI signals --- using the NK model does not reduce the size of the parameter space or the number of waveforms required to cover it. For a given set of parameters, however, the model is much simpler to evaluate than a perturbative waveform and that is where the computational savings for data analysis arise. Construction of an NK waveform is done in two stages --- (i) generation of the trajectory for an object inspiralling through a sequence of quasi-geodesic orbits; (ii) construction of an approximate waveform for an object moving on this trajectory.

\subsection{Inspiral Trajectory Generation}
To construct the inspiral trajectory, we first compute the phase space evolution of the object, i.e., how the energy ($E$), angular momentum ($L_z$) and Carter constant ($Q$) of the orbit evolve with time. This is done by deriving suitable expressions for ${\rm d}E/{\rm d}t$ etc. as a function of the orbital parameters, and then integrating them through phase space. The expressions we use are built on second order post-Newtonian (PN) results\cite{ghk,GG06}. Using 2PN results directly leads to pathological behavior for nearly circular orbits, but this can be corrected by amending the circular pieces of the fluxes\cite{GG06}. The trajectories can be further improved by using fits to data derived from perturbation theory, i.e., based on solution of the Teukolsky equation. We have done this for circular orbits of arbitrary inclination, but not yet for generic orbits since perturbative data for such situations is only now becoming available\cite{drasco06}. The resulting phase space evolution equations are given in detail in Gair \& Glampedakis 2006\cite{GG06}. For circular inclined inspirals, these fluxes agree with perturbative results to an accuracy of $1\%$ for orbits with periapse greater than $\sim 5M$, and to an accuracy of $<5\%$ for orbits with periapse greater than $\sim2M$. For eccentric orbits, the fluxes agree to $\sim 5\%$ for orbits with periapse greater than $5M$, but this increases to a few tens of percent for orbits that come very close to the central black hole. 

Once the phase space trajectory ($E(t)$, $L_z(t)$, $Q(t)$) has been obtained, the inspiral trajectory is derived by integrating the Kerr geodesic equations ${\rm d}r/{\rm d}t = R(r,\theta,E,L_z,Q)$, ${\rm d}\theta/{\rm d}t = \Theta(r,\theta,E,L_z,Q)$, ${\rm d}\phi/{\rm d}t = \Phi(r,\theta,E,L_z,Q)$, 
with the time-dependent $E$, $L_z$ and $Q$ inserted on the right hand side. We thus obtain the particle trajectory in Boyer-Lindquist coordinates, $(r(t), \theta(t), \phi(t))$.
  
\subsection{Waveform Construction}
After computing the particle trajectory in Boyer-Lindquist coordinates, we may construct a corresponding trajectory in a pseudo-flat space by identifying these coordinates with spherical polar coordinates. A waveform can then be obtained by supposing that there was a particle moving on such a trajectory in flat space, and using a weak-field GW emission formula. This approach is inconsistent in the sense that it neglects the stress-energy that is causing the particle to move on the trajectory, but it appears {\it post facto} to work well. We have constructed waveforms using the Press formula\cite{press77} (valid for weak-field, fast motion sources) and also using the quadrupole and quadrupole-octupole formulae obtained by expanding the Press expression in $v/c$. Based on a balance between ease of computation and accuracy, it appears that the quadrupole-octupole formula is optimal. This waveform construction is described in more detail in Babak et al. 2006\cite{babak06}.

\section{Application to LISA}
For both generic geodesic orbits, and for circular inclined inspiral orbits, the overlap between the NK waveforms and more accurate adiabatic waveforms is very high. For orbits with periapse greater than $\sim 5M$ the overlaps are typically greater than $95\%$, but this degrades for orbits that come deep into the strong field near the black hole\cite{babak06}. The waveforms are sufficiently cheap to be generated in the large numbers required for LISA data analysis, while their high faithfulness suggests that they will also be able to constrain the source parameters quite well. NK waveforms are already being used for scoping out LISA data analysis\cite{gair04}, and their high accuracy indicates that they could play an important role in source detection for LISA, and quite possibly for parameter estimation as the first stage of a hierarchical search.

The NK waveforms can be further improved in several ways --- (i) inclusion of PN conservative self-force corrections, i.e., the piece of the self-force that does not dissipate. We have already demonstrated how this can be done to lowest order for the simple case of circular inspirals in the Schwarzschild spacetime\cite{babak06}. Inclusion of this effect will provide information on the relative influence of conservative corrections on the phasing of generic EMRI waveforms, currently a matter of some debate. (ii) Addition of ``tail terms'', i.e., the effect of radiation back-scattering off the background geometry. This can be done by expanding the Teukolsky function and should help to improve the accuracy of the NK waveforms for strong-field orbits. (iii) Improvement of the flux expressions, i.e., ${\rm d}E/{\rm d}t$ etc., for eccentric orbits by using fits to perturbative data. This will ensure the NK waveforms can match true EMRI signals for longer segments of the inspiral. These three improvements will be implemented in the future to further develop this model as a tool for data analysis.

\section*{Acknowledgments}
The work described in this paper was done in collaboration with Stanislav Babak, Hua Fang, Kostas Glampedakis and Scott Hughes\cite{GG06,babak06}. JG's work was supported by St.Catharine's College, Cambridge.

\end{document}